\documentclass[trackchanges,twocolumn]{aastex701}
\usepackage{amsmath}
\usepackage{comment}

\newcommand{\new}[1]{{#1}}
\newcommand{\neww}[1]{{#1}}
\begin{document}

%\title{Gravitational-wave spin measurements reveal that most binary black holes do not merge alone}
%\title{Near-perpendicular spin-orbit angles in the detected gravitational-wave\\population of binary black hole mergers}
%\title{In-plane Black-hole Spin Measurements Suggest\\Most Gravitational-wave Mergers Form in Triples}
\title{Gravitational-wave Observations Suggest Most Black Hole Mergers Form in Triples}

\author[orcid=0000-0003-2340-8140]{Jakob Stegmann}
\affiliation{Max Planck Institute for Astrophysics, Karl-Schwarzschild-Str. 1, 85741 Garching, Germany}
\email[show]{jstegmann@mpa-garching.mpg.de}  

\author[orcid=0000-0003-3138-6199]{Fabio Antonini} 
\affiliation{Gravity Exploration Institute, School of Physics and Astronomy,
Cardiff University, Cardiff, CF24 3AA, UK}
\email[]{antoninif@cardiff.ac.uk}

\author[orcid=0000-0002-6105-6492]{Aleksandra Olejak}
\affiliation{Max Planck Institute for Astrophysics, Karl-Schwarzschild-Str. 1, 85741 Garching, Germany}
\email[]{aolejak@mpa-garching.mpg.de} 

\author[orcid=0000-0001-7616-7366]{Sylvia Biscoveanu}
\affiliation{Department of Physics, Princeton University, Princeton, NJ 08544, USA}
\email[]{sbisco@princeton.edu} 

\author{Vivien Raymond} 
\affiliation{Gravity Exploration Institute, School of Physics and Astronomy,
Cardiff University, Cardiff, CF24 3AA, UK}
\email[]{RaymondV@cardiff.ac.uk}

\author[orcid=0000-0001-5799-4155]{Stefano Rinaldi}
\affiliation{Institut für Theoretische Astrophysik, ZAH, Universität Heidelberg, Albert-Ueberle-Str.~2, 69120 Heidelberg, Germany}
\email[]{stefano.rinaldi@uni-heidelberg.de} 

\author{Elizabeth Flanagan} 
\affiliation{Gravity Exploration Institute, School of Physics and Astronomy,
Cardiff University, Cardiff, CF24 3AA, UK}
\email[]{}

%% Use the \collaboration command to identify collaborations. This command
%% takes an optional argument that is either a number or the word "all"
%% which tells the compiler how many of the authors above the command to
%% show. For example "\collaboration[all]{(DELVE Collaboration)}" wil include
%% all the authors above this command.
%%
%% Mark off the abstract in the ``abstract'' environment. 
\begin{abstract}
The spin-orbit tilt angles $\theta_{1(2)}$ of merging stellar-mass black holes provide key insights into their astrophysical origin. \new{Non-parametric population modelling of \citet{GWTC4pop} shows} that the spin-orbit tilt distribution of mergers in the latest Gravitational-Wave Transient Catalog 4.0 exhibits a global peak at near-perpendicular directions $\cos\theta_{1(2)}\approx0$. Here, we recover this feature using hierarchical Bayesian inference with parametric models that are tailored to enhance the diagnostic power about astrophysical formation channels. We find that the spin distribution of the low-mass bulk of the binary black hole merger population ($m_1\lesssim 44.3^{+8.7}_{-4.6}\,\rm M_\odot$) can be well-modelled by a dominant Gaussian component that peaks at $\cos\theta_{1(2)}\approx0$, possibly mixed with a subdominant isotropic component. Models that include a component with spins preferentially aligned with the orbit are disfavoured by current data (with Bayes factors $|\Delta\ln\mathcal{B}|\approx1$ to~$3$) and constrain its contribution to be \neww{likely} small ($\xi\sim\mathcal{O}(1)\,\%$), \neww{although large contributions cannot yet ruled out with certainty}.
If these findings are reinforced by more detections, they would challenge any major contribution from the traditional isolated-binary formation scenario yielding closely aligned spins. Instead, the dominant component with near-perpendicular spins matches expectations
from the evolution of isolated massive stellar triples in the galactic field, where the Lidov--Kozai effect naturally produces a unique overabundance of mergers with  $\cos\theta_{1(2)}\approx0$.
\end{abstract}

%% Keywords should appear after the \end{abstract} command. 
%% The AAS Journals now uses Unified Astronomy Thesaurus (UAT) concepts:
%% https://astrothesaurus.org
%% You will be asked to selected these concepts during the submission process
%% but this old "keyword" functionality is maintained in case authors want
%% to include these concepts in their preprints.
%%
%% You can use the \uat command to link your UAT concepts back its source.

%% From the front matter, we move on to the body of the paper.
%% Sections are demarcated by \section and \subsection, respectively.
%% Observe the use of the LaTeX \label
%% command after the \subsection to give a symbolic KEY to the
%% subsection for cross-referencing in a \ref command.
%% You can use LaTeX's \ref and \label commands to keep track of
%% cross-references to sections, equations, tables, and figures.
%% That way, if you change the order of any elements, LaTeX will
%% automatically renumber them.

\section{Introduction}
A decade after the first direct detection of gravitational waves from merging binary black holes \citep{Abbott2016}, the observational sample detected by the LIGO-Virgo-KAGRA (LVK) interferometers has grown to about two hundred of these events \citep{2015CQGra..32g4001L, 2015CQGra..32b4001A, 2013PhRvD..88d3007A, 2012CQGra..29l4007S,2018LRR....21....3A,10.1093/ptep/ptaa125,GWTC4Updating}. Yet, a central question remains open: What are the formation mechanisms behind the observed mergers of black holes? 

Resolving this question has become a major challenge in gravitational-wave astronomy as most proposed binary black hole formation channels---such as those arising from isolated binary-star evolution \citep[][]{2016ApJ...819..108B}, active galactic nuclei \citep[][]{Stone2017}, or ultra-wide binaries perturbed by the Galaxy \citep{2024ApJ...972L..19S}---are notoriously difficult to model in a predictive way \cite[e.g.,][]{2018PhRvD..98h4036G,Baibhav2024}. Their outcomes depend on numerous uncertain assumptions, making their predictions highly flexible; with sufficient tuning,  they can be made to reproduce a wide range of features in gravitational-wave observations.

Among all proposed formation channels, the evolution of triples---which are by far the most abundant observed configuration of massive black hole progenitor stars \citep{Moe2017,Offner2023}---stands out by making a unique, testable prediction for the spin orientation of merging black holes. In this scenario, the gravitational perturbation from a distant companion drives large-amplitude eccentricity oscillations (the ``Lidov--Kozai'' effect) of black holes formed in the inner binary \citep[][]{Zeipel1910,Lidov1962,Kozai1962,Naoz2016}. Efficient gravitational-wave emission during close pericentre passages can then lead the binary to inspiral and merge \citep{Silsbee2017,Antonini2017,Grishin2018,Liu2018,Antonini2018,Rodriguez2018,Mangipudi2022,Stegmann2022,2025A&A...699A.272V,Dorozsmai2025,Stegmann2025}. Under well-defined conditions, the combined action of Lidov--Kozai oscillations, gravitational-wave emission, and relativistic spin precession drives the component spins to flip into the orbital plane \new{\citep{Antonini2018,Liu2018,Rodriguez2018,Su2021}}, producing an overall excess of systems with spin--orbit misalignments near perpendicular (cf. Section~\ref{sec:astro}). This configuration is so unusual that other formation channels typically require additional, ad hoc assumptions to reproduce it \citep[e.g.,][]{StegmannAntonini,Tauris2022,Baibhav2024,Vaccaro2024}. The triple scenario therefore provides a robust formation pathway with a clear, falsifiable prediction.

Tentative evidence for a global peak in
the binary black hole population at  $\cos\theta\approx0$—corresponding to a spin--orbit tilt angle of about $90^{\circ}$—was identified in analyses of earlier LVK catalogues \new{through parametric modelling in \citet{Vitale2022} and was subsequently corroborated with more agnostic spin models \citep{2023ApJ...946...16E,2023PhRvD.108j3009G}.} Support for this feature has strengthened with the Gravitational-Wave Transient Catalog 4.0 \citep[GWTC-4.0;][see Figure~7 in the latter]{GWTC2data,GWTC3data,GWTC4data,GWTC4Updating,GWTC4pop}, particularly in the non-parametric (``weakly modelled'') B-spline population model~\citep{2023ApJ...946...16E}, which makes minimal \textit{a priori} assumptions, but can be difficult to interpret astrophysically. The LVK's  parametric (``strongly modelled'') default spin-tilt model~\citep{GWTC4pop} also recovers weak evidence for this peak. Here we aim to assess and interpret this feature using a more astrophysically motivated parametric framework, that allows direct comparison with predictions from formation channels.

\section{Methods}\label{sec:methods}
While many studies have shown that the population spin--tilt distribution is a powerful discriminator of black hole formation channels~\citep{2017CQGra..34cLT01V, 2017MNRAS.471.2801S, 2017PhRvD..96b3012T, Vitale2022}, tilt measurements for individual events carry large uncertainties~\citep{2014PhRvL.112y1101V, 2020PhRvR...2d3096P, 2021PhRvD.104j3018B}, limiting population-level inference~\citep{2024PhRvD.109j4036M, VitaleMould2025}. Population constraints have therefore often relied on the effective spin
$
\chi_{\rm eff}={\chi_1\cos\theta_1+q\,\chi_2\cos\theta_2}/\new{(1+q)},
$
where $\chi_{1(2)}$ and $\theta_{1(2)}$ denote the component spin magnitudes and tilt angles, respectively, and $0<q\le1$ is the binary mass ratio~\citep{Damour2001}. While $\chi_{\rm eff}$ is more precisely measured, its inferred population distribution peaks near zero~\citep{2020ApJ...895..128M, 2021PhRvD.104h3010R, Banagiri2025b, GWTC4pop}. This is intrinsically ambiguous: it can arise from small spin magnitudes  or from substantial spin-orbit misalignment, and it is further compounded by the mass-weighted combination of both components. This motivates investigating the population distribution of individual spin magnitudes and tilts directly, rather than relying on $\chi_{\rm eff}$ alone.
In what follows, we therefore employ a hierarchical Bayesian inference method \citep[e.g.,][]{Mandel2019} to infer hyper-parameters describing the component spin properties of the binary black hole merger population.

We use the \texttt{GWPopulation} code \citep[][]{Talbot2019,Talbot2025} and the public detector sensitivity estimates \citep{2025PhRvD.112j2001E, GWTCsensitivity} and individual event posteriors from GWTC-4.0 \citep{GWTC2data,GWTC3data,GWTC4data}. \new{In using \texttt{GWPopulation}, we limit the variance of the Monte Carlo–estimated log-likelihood to be below 1} and use the posterior samples obtained with the \texttt{NRSur7dq4} waveform model~\citep{2019PhRvR...1c3015V} where available for new events in GWTC-4.0, the mixed-waveform samples otherwise, and the \texttt{IMRPhenomXPHM} waveform samples~\citep{2021PhRvD.103j4056P, 2025PhRvD.111j4019C} for all events that appear in previous catalogs. Our method is exemplified in a \new{public script\footnote{\url{https://github.com/stegmaja/black-hole-spin-orbit-tilts}}} accompanying this work and detailed in the following.

%We define $\chi_{1(2)}$ as the spin magnitude of the primary (secondary) black hole and $\cos\theta_{1(2)}=\boldsymbol{\hat\chi}_{1(2)}\cdot\boldsymbol{\hat L}$ as its tilt angle with respect to the binary orbital angular momentum $\boldsymbol{\hat{L}}$. Furthermore,  we denote the mass of the primary (secondary) as $m_{1(2)}$. To streamline the notation, 

\new{In order to streamline our notation,  we introduce two helper functions}
\begin{align}
    \pi_\chi(\chi_i|\mu_\chi,\sigma_\chi) =\,& \mathcal{N}_{[0,1]}(\chi_1 | \mu_\chi, \sigma_\chi)\mathcal{N}_{[0,1]}(\chi_2 | \mu_\chi, \sigma_\chi),\label{eq:pi_chi} \\
    \pi_t(\cos\theta_i|\mu_t,\sigma_t)
    =\,&\mathcal{N}_{[-1,1]}(\cos\theta_1 | \mu_t, \sigma_t)\nonumber\\
    &\times\mathcal{N}_{[-1,1]}(\cos\theta_2 | \mu_t, \sigma_t),
    \label{eq:pi_cos} 
\end{align}
where $\mathcal{N}_{[a,b]}(x|\mu, \sigma)$ is a truncated Gaussian distribution within $a\le x\le b$ and $\mu$ and $\sigma$ are its mean and standard deviation, respectively. Using Equations~\eqref{eq:pi_chi} and~\eqref{eq:pi_cos} we \new{define the distribution function}
%\onecolumngrid
%\begin{equation}
%    \pi=(1-\zeta)\left[\xi\,\pi_t(\cos\theta_i|\mu_t,\sigma_t)\,\pi_\chi(\chi_i|\mu_\chi,\sigma_\chi)+\frac{1-\xi}{4}\,\pi_\chi(\chi_i|\mu^{\rm Iso}_\chi,\sigma^{\rm Iso}_\chi)\right]+\frac{\zeta}{4}\,\pi_\chi(\chi_i|\mu^{\rm HighIso}_\chi,\sigma^{\rm HighIso}_\chi)\label{eq:pi},
%\end{equation}
%\twocolumngrid
\begin{align}
    &\new{\pi(\chi_i,\cos\theta_i|\Lambda)}%\nonumber\\
    =(1-\zeta)\Big[\xi\,\pi_t(\cos\theta_i|\mu_t,\sigma_t)\pi_\chi(\chi_i|\mu_\chi,\sigma_\chi)\nonumber\\
    &+\frac{1-\xi}{4}\,\pi_\chi(\chi_i|\mu^{\rm Iso}_\chi,\sigma^{\rm Iso}_\chi)\Big]+\frac{\zeta}{4}\,\pi_\chi(\chi_i|\mu^{\rm HighIso}_\chi,\sigma^{\rm HighIso}_\chi)\label{eq:pi},
    \\ &\new{\{\zeta,\xi,\mu_t,\sigma_t,\mu_\chi,\sigma_\chi,\mu^{\rm Iso}_\chi,\sigma^{\rm Iso}_\chi,\mu^{\rm HighIso}_\chi,\sigma^{\rm HighIso}_\chi\} \in \Lambda} \nonumber
\end{align}
\begin{figure}
    \centering
    \includegraphics[width=\linewidth]{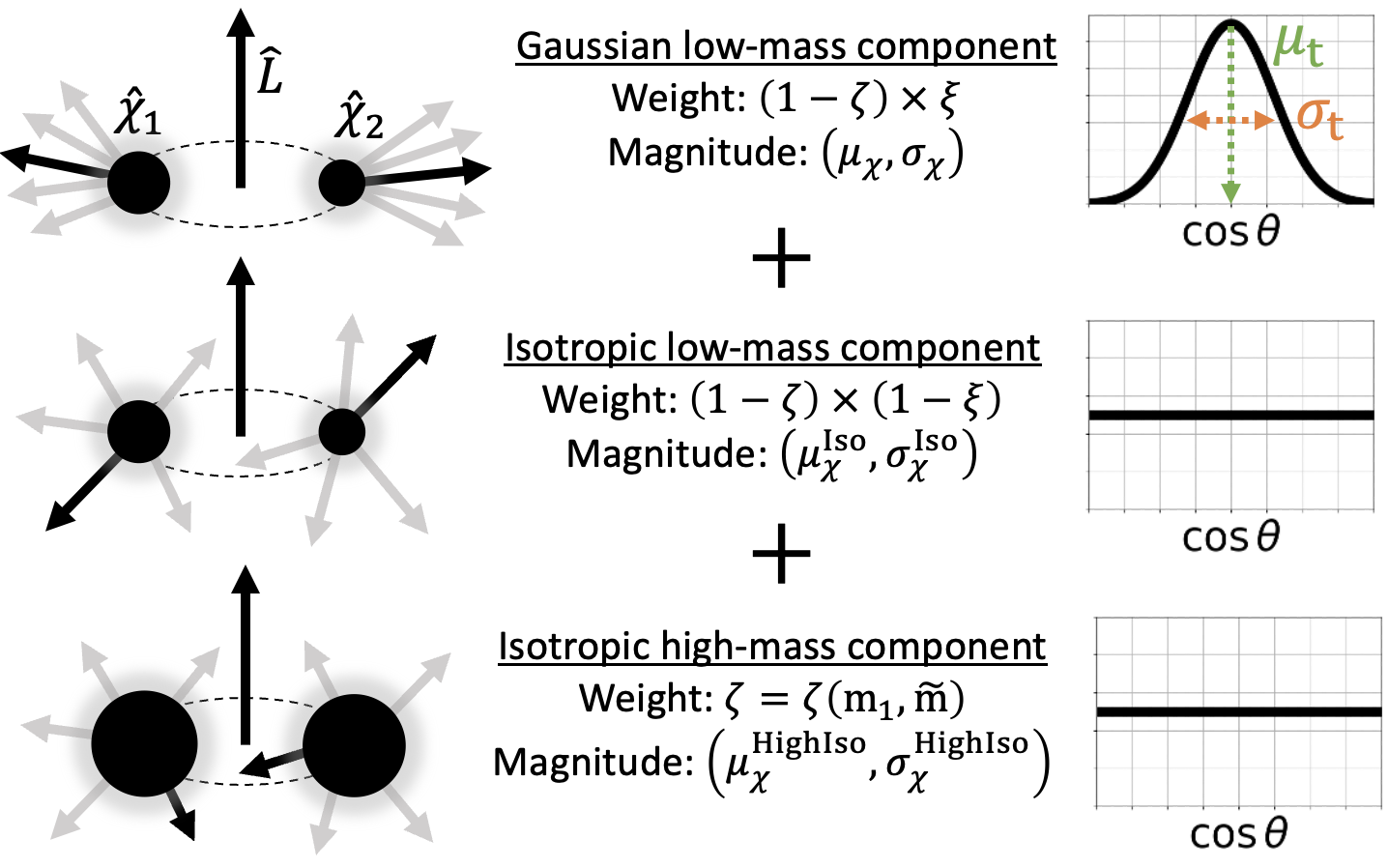}
    \caption{Schematic overview of the parametric spin population model described in Equation~\eqref{eq:pi}. Below a mass cut-off $m_1\lesssim\tilde{m}$ the population is described by a mixture between a component whose spin directions $\cos\theta_{1(2)}=\boldsymbol{\hat\chi}_{1(2)}\cdot\boldsymbol{\hat L}$ follow a truncated Gaussian distribution (mixing fraction $\xi$) and a component with isotropic spin directions (mixing fraction $1-\xi$). Above $m_1\gtrsim\tilde{m}$ the spins also follow an isotropic distribution. Each of the three components are allowed to follow different spin magnitude distributions.}
    \label{fig:sketch}
\end{figure}
\begin{table*}[]
    \centering
    \begin{tabular}{c|cccccc|ccc}
         Models & \multicolumn{6}{c|}{Definitions} & $Y_{\rm vs.~aligned}$ & $Y_{\rm vs.~anti-aligned}$ & $\Delta\ln\mathcal{B}$ \\
         \hline
         (1) & $\zeta=0$ & $\xi=1$ & $-1\le\mu_t\le1$ & $\sigma_t\sim\mathcal{N}_{[0.1,4]}$ & & & $3.0^{+3.6}_{-1.5}$ & $10.8^{+23.9}_{-6.7}$ & $-10.1$ \\
         (2) & $\zeta=0$ & $\xi=1$ & $\mu_t=1$ & $\sigma_t\sim\mathcal{N}_{[0.1,4]}$ & & & $0.7^{+0.1}_{-0.1}$ & $2.5^{+0.7}_{-0.5}$ & $-13.2$ \\
         (3) & $\zeta=0$ & $0\le\xi\le1$ & $-1\le\mu_t\le1$ & $\sigma_t\sim\mathcal{N}_{[0.1,4]}$ & & & $1.4^{+1.5}_{-0.4}$ & $2.7^{+3.3}_{-1.5}$ & $-6.3$ \\
         (4) & $\zeta=0$ & $0\le\xi\le1$ & $-1\le\mu_t\le1$ & $\sigma_t\sim\mathcal{U}_{[0.1,4]}$ & & & $1.0^{+0.2}_{-0.1}$ & $1.2^{+1.0}_{-0.2}$ & $-6.9$ \\
         (5) & $\zeta=0$ & $0\le\xi\le1$ & $\mu_t=1$ & $\sigma_t\sim\mathcal{N}_{[0.1,4]}$ & & & $0.8^{+0.1}_{-0.1}$ & $1.1^{+0.1}_{-0.1}$ & $-7.2$ \\
         (6) & $\zeta=0$ & $0\le\xi\le1$ & $-1\le\mu_t\le1$ & $\sigma_t\sim\mathcal{N}_{[0.1,4]}$ & $\mu_\chi=\mu_\chi^{\rm Iso}$ & $\sigma_\chi=\sigma_\chi^{\rm Iso}$ & $2.5^{+2.5}_{-1.2}$ & $4.5^{+5.7}_{-2.3}$ & $-11.2$ \\
         (7) & $\zeta=0$ & $0\le\xi\le1$ & $-1\le\mu_t\le1$ & $\sigma_t\sim\mathcal{U}_{[0.1,4]}$ & $\mu_\chi=\mu_\chi^{\rm Iso}$ & $\sigma_\chi=\sigma_\chi^{\rm Iso}$ & $1.2^{+2.0}_{-0.3}$ & $2.3^{+4.2}_{-1.1}$ & $-14.7$ \\
         (8) & $\zeta(m_1,\tilde{m})$ & $0\le\xi\le1$ & $\mu_t=1$ & $\sigma_t\sim\mathcal{N}_{[0.1,4]}$ & & & $0.9^{+0.1}_{-0.1}$ & $1.1^{+0.7}_{-0.1}$ & $-1.6$ \\
         (9) & $\zeta(m_1,\tilde{m})$ & $0\le\xi\le1$ & $-1\le\mu_t\le1$ & $\sigma_t\sim\mathcal{N}_{[0.1,4]}$ & & & $1.7^{+1.9}_{-0.7}$ & $3.4^{+4.7}_{-1.9}$ & $0.0$ \\
         \multicolumn{10}{l}{Models: (1) \texttt{Gaussian} (2) \texttt{Aligned} (3) \texttt{Gaussian + Isotropic} (4) \texttt{Uniform Gaussian + Isotropic} (5) \texttt{Aligned + Isotropic}}\\
         \multicolumn{10}{l}{(6) \texttt{LVK Gaussian + Isotropic} (7) \texttt{LVK Uniform Gaussian + Isotropic} (8) \texttt{Aligned + Isotropic + Cut}}\\
         \multicolumn{10}{l}{(9) \texttt{Gaussian + Isotropic + Cut}}\\
    \end{tabular}
    
    \caption{Overview of parametric spin models studied in this work. The truncated normal distribution $\mathcal{N}_{[0.1,4]}$ used as a prior for $\sigma_t$ in most models is assuming a mean and standard deviation of $\mu=0$ and $\sigma=1/2$, respectively. The two \texttt{LVK} models enforce that the Gaussian and isotropic components are following the same spin magnitude distribution, i.e., $\mu_\chi^{\rm Iso}$ and $\sigma_\chi^{\rm Iso}$ in Equation~\eqref{eq:pi} are replaced by $\mu_\chi$ and $\sigma_\chi$, respectively. Other model specifications are detailed in Section~\ref{sec:methods}. The last three columns contain the excess fractions defined in Equations~\eqref{eq:excess1} and~\eqref{eq:excess2} for $\delta=0.1$ and resulting Bayes factors relative to the \texttt{Gaussian + Isotropic + Cut} model, which are discussed in Section~\ref{sec:results}.}
    \label{tab:models}
\end{table*}
where $0\le\xi\le1$ is a mixing fraction describing the Gaussian component (while $1-\xi$ is isotropic), $\zeta=\zeta(m_1;\tilde{m})$ is defined below and describes a mass-dependent transition to a fully isotropic component above $m_1\gtrsim\tilde{m}$, and all $\mu$'s and $\sigma$'s define means and standard deviations of the truncated normal distributions in Equations~\eqref{eq:pi_chi} and~\eqref{eq:pi_cos}, respectively. In Figure~\ref{fig:sketch}, we provide a schematic overview of the various components involved in Equation~\eqref{eq:pi}. \new{We summarise our modelling of the black hole mass and redshift distributions in Appendix~\ref{appendix-1}.}

Equation~\eqref{eq:pi} is used to investigate population models that reflect merger contributions from various astrophysical formation channels. An isotropic component is expected from binary black hole mergers formed through close few-body encounters in dense environments such as star clusters \citep[e.g.,][]{Rodriguez2016}. %possibly with a small fraction whose spins realign due to stellar collisions and accretion of collision debris \citep{Fulya2025}.
Mergers from the evolution of isolated binary and triple systems are expected to have preferentially aligned \citep{Kalogera2000,Farr2017,2018PhRvD..98h4036G,Belczynski2020,Bavera2020,OlejakSpins2021,Broekgaarden2022} or near-perpendicular spin-orbit orientations \citep{Antonini2018,Liu2018}, respectively, and can be reflected by the Gaussian component with appropriate $\mu_t$ and $\sigma_t$ (cf.~Section~\ref{sec:astro}). Meanwhile, mergers assembled in active galactic nucleus disks are less certain but generally expected to inherit a preferred axis set by the disk angular momentum \citep{Yang2019a,Vaccaro2024}.

Thus, we introduce a set of parametric models whose properties are summarised in Table~\ref{tab:models} and priors on the population parameters are defined at the end of this section. In the \texttt{Gaussian} and \texttt{Aligned} models we assume that the spin-orbit tilts can be described by a single truncated Gaussian whose location is allowed to be free within $-1\le\mu_t\le1$ or enforced at alignment $\mu_t=1$, respectively. In these single-component models we impose $\zeta=0$ and $\xi=1$. In the next set of models \texttt{Gaussian + Isotropic}, \texttt{Uniform Gaussian + Isotropic}, and \texttt{Aligned + Isotropic} we allow for a free mixing with an isotropic component by allowing $0\le\xi\le1$. In these models, the Gaussian and isotropic components are allowed to follow different spin magnitude distributions each parametrised by truncated Gaussians with $(\mu_\chi,\sigma_\chi)$ and $(\mu_\chi^{\rm Iso},\sigma_\chi^{\rm Iso})$, respectively. If both components are thought to represent two different astrophysical subpopulations, e.g., from the evolution of isolated stellar few-body systems (binaries, triples, etc.) and from few-body encounters in dense environments (e.g., globular clusters), respectively, there is \textit{a priori} no reason to assume they should follow the same spin magnitude distribution. This is different from the default modelling of \citet{GWTC4pop}, which enforces the same spin magnitude distribution in both components. To test the consequences of this assumption, we include two models \texttt{LVK Gaussian + Isotropic} and \texttt{LVK Uniform Gaussian + Isotropic}, where we replace $(\mu_\chi^{\rm Iso},\sigma_\chi^{\rm Iso})$ by $(\mu_\chi,\sigma_\chi)$. 

Finally, \citet{Antonini2025,2025PhRvL.134a1401A} identify a transition to an isotropic high-spin population for primary masses above a mass threshold $\tilde{m}=45^{+6}_{-4}M_\odot$, 
%which coincides with a sparsity of secondary masses inferred by \citet{Tong2025}, 
which they
interpret 
as evidence for a high-mass tail produced through hierarchical black hole mergers \citep[e.g.,][]{Rodriguez2019,GerosaFishbach}.  
Consistently, \citet{2022ApJ...941L..39W} identify a similar transition to a higher-spin population at ${46.1}_{-5.1}^{+5.6}\,{M}_{\odot }$ based on GWTC--3 data.
%, which coincides with a sparsity of secondary masses inferred by \citet{Tong2025}. It is an active debate \citep{Antonini2025,Tong2025,Ray2025,Tong2025b,Wang2025,Banagiri2025} whether or not this indicates the onset of the upper mass gap where black holes are not expected to form through standard stellar collapse \citep{Bond1984,Heger2002,Fryer2001} but could be populated through hierarchical black hole mergers \citep{Rodriguez2019,Yang2019a,Antonini2019,ArcaSedda2020,Mapelli2021,AntoniniGieles2023,Chattopadhyay2023,Fragione2023,Vaccaro2024,Torniamenti2024,Mahapatra2025,StegmannOlejak2025,Gilbaum2025} or stellar collisions in dense environments  \citep[e.g.,][]{Kremer2020,Costa2022}. Here, we remain agnostic about the nature of the high-mass population but 
Here, we focus on the lower-mass bulk of the astrophysical merger population, but ensure that our analysis is not contaminated by the  distinct spin properties of the high-mass population. Therefore, 
we introduce a \texttt{Gaussian + Isotropic + Cut} model, in which
\begin{equation}\label{eq:zeta}
    \new{\zeta(m_1,\tilde{m})=\frac{1}{1+\exp[\tilde{k}(\tilde{m}-m_1)]}}
\end{equation}
is a sigmoid function that smoothly transitions from the \texttt{Gaussian + Isotropic} model at low primary masses $m_1\lesssim\tilde{m}$ to an isotropic component with separate spin magnitude distribution $(\mu_\chi^{\rm HighIso},\sigma_\chi^{\rm HighIso})$ at high masses $m_1\gtrsim\tilde{m}$. Another model \texttt{Aligned + Isotropic + Cut} additionally enforces $\mu_t=1$. \new{As a default, we fix $\tilde{k}=1$ in all our models which incorporate the mass cut. We have also tested one variant of the \texttt{Gaussian + Isotropic + Cut} model, where we allow $\tilde{k}$ to be a free hyper-parameter of the model between 0.1 and 2.0 (where larger values of $\tilde{k}$ lead to a sharper transition in mass). As a result, we found that $\tilde{k}=0.9^{+0.7}_{-0.6}$ is poorly constrained with a posterior distribution which is largely consistent with the prior on $\tilde{k}$, which we assumed to be uniform. \neww{As shown in Appendix~\ref{appendix-2}, }we also found all other posterior distributions of the model parameters to be largely similar between the runs with free and a fixed $\tilde{k}=1$.}

\begin{figure*}
    \centering
    \includegraphics[width=\linewidth]{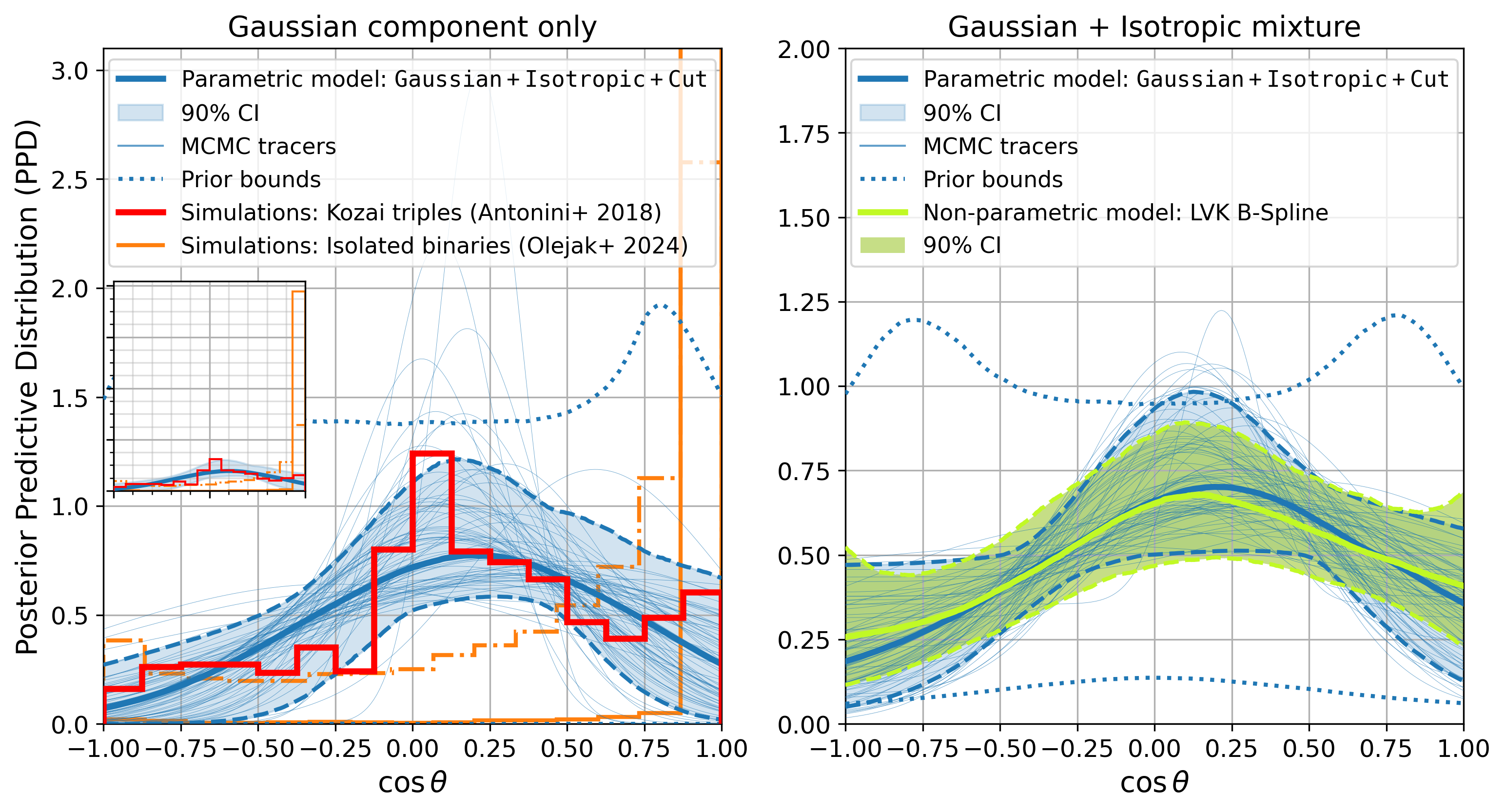}
    \caption{Posterior predictive distribution (PPD) of the black hole spin tilts of the low-mass population ($m_1\lesssim\tilde{m}= 44.3^{+8.6}_{-4.6}\,\rm M_\odot$) in our parametric \texttt{Gaussian + Isotropic + Cut} model (blue). The PPD is constructed by taking the quantiles (median indicated by dashed lines, $90\,\%$ interval by shaded envelope) across MCMC samples (thin lines). The left panel shows the PPD of only the low-mass Gaussian component, the right panel includes the low-mass isotropic component. The green colour shows the non-parametric B-Spline model of \citet{GWTC4pop,GWTC4pop-zenodo}. \new{In both panels, the dotted blue lines indicate the prior bounds ($90\,\%$), i.e., they show the PPDs if all hyperparameters are drawn from their priors.} The red line shows simulation outcomes of binary black hole mergers which are caused by the Lidov--Kozai effect in hierarchical triples \citep{Antonini2018}. The orange lines show simulation outcomes of binary black hole mergers from isolated binary star evolution \citep{Olejak2024} assuming high natal kicks at black hole formation that are drawn from a Maxwellian velocity distribution with $\sigma=133\,\rm km\,s^{-1}$ (dash-dotted) or natal kicks lowered by fallback \citep{Fryer2012} with $\sigma=265\,\rm km\,s^{-1}$ (solid). Since we cut the y-axis, the small inset (linear axes) shows that the orange models are strongly concentrated at $\cos\approx1$. In the right panel, the binary black hole sketch depicts a merger with $\cos\theta_{1(2)}\approx0$, where, for visualisation purposes, we pick near-opposite in-plane directions.}
    \label{fig:PPD-tilts}
\end{figure*}

%Regarding the priors for the hyper-parameters of the spin distribution in all models, 
We assume uniform distributions, $\mathcal{U}_{[a,b]}$ between $a$ and $b$,  for the priors of most model hyper-parameters: 
\begin{itemize}
    \item $\mathcal{U}_{[0,1]}$ for $\mu_\chi$, $\mu_\chi^{\rm Iso}$, $\mu_\chi^{\rm HighIso}$, and $\xi$,
    \item $\mathcal{U}_{[0.1,1]}$ for $\sigma_\chi$, $\sigma_\chi^{\rm Iso}$, and $\sigma_\chi^{\rm HighIso}$,
    \item $\mathcal{U}_{[10,100]}$ for $\tilde{m}/\rm M_\odot$.
\end{itemize} 
In the \texttt{Uniform Gaussian + Isotropic} and \texttt{LVK Uniform Gaussian + Isotropic} models we also assume a uniform prior $\mathcal{U}_{[0.1,4]}$ for $\sigma_t$, similar to population models of \citet{GWTC4pop}. 
In all other models 
we assume $\mathcal{N}_{[0.1,4]}(\mu=0,\sigma=1/2)$, as wide spin tilt distributions (e.g., reflected in a uniform prior up to $\sigma_t=4$) are inconsistent with most astrophysical models, which instead tend to exhibit more pronounced peaks and are better modelled by a width of $\sigma\approx1/2$ or less \citep[e.g.,][and astrophysical models in Figure~\ref{fig:PPD-tilts} below]{Baibhav2024}. 

\section{Results}\label{sec:results}
At first, we quantify the statistical significance of the global peak at $\cos\theta\approx0$ in the non-parametric B-Spline model of \citet{GWTC4pop}. We follow a similar approach as \citet{Vitale2022} and define
\begin{align}
    Y_{\rm vs.~aligned}(\delta)&=\frac{p(\cos\theta\in[-\delta/2,\delta/2])}{p(\cos\theta\in[1-\delta,1])},\label{eq:excess1}\\
    Y_{\rm vs.~anti-aligned}(\delta)&=\frac{p(\cos\theta\in[-\delta/2,\delta/2])}{p(\cos\theta\in[-1,-1+\delta])},\label{eq:excess2}
\end{align}
to quantify the excess of near-perpendicular spin-orbit angles within $\pm\,\delta/2$ compared to near-aligned ($\cos\theta\in[1-\delta,1]$) and near-anti-aligned configurations ($\cos\theta\in[-1,-1+\delta]$), respectively. For the public hyper-parameter posterior samples of the B-Spline model \citep{GWTC4pop,GWTC4pop-zenodo} and $\delta=0.1$ we find $Y_{\rm vs.~aligned}=1.5^{+0.7}_{-0.5}$ and $Y_{\rm vs.~anti-aligned}=2.5^{+1.8}_{-1.0}$, which suggest a statistically significant excess at $\cos\theta\approx0$ and a skewness to positive values of $\cos\theta$.

Concerning our parametric models, we find that this spin distribution is best represented by \texttt{Gaussian + Isotropic + Cut} model which outperforms all other models, but \texttt{Aligned + Isotropic + Cut}, strongly \citep{Jeffreys,Kass} with Bayes factors of $|\Delta\ln\mathcal{B}|>6.3$ (summarised in the last column of Table~\ref{tab:models}). \new{Evidences for each model are calculated using the \texttt{Dynesty} nested sampler~\citep{2020MNRAS.493.3132S}, with error estimates on the order of $\mathcal{O}(0.1)$ in log-evidence, much smaller than the log Bayes factors we calculate in favour of the \texttt{Gaussian + Isotropic + Cut} model.}
This aligns with the  astrophysical expectation of an isotropic component from dense environments above some mass threshold and a combination with mergers from stellar evolution processes below the threshold \citep[e.g.,][]{Rodriguez2019,GerosaFishbach}. Allowing for a free Gaussian is always statistically preferred over enforced spin-orbit alignment. Evidence of \texttt{Gaussian + Isotropic + Cut} against the aligned counterpart \texttt{Aligned + Isotropic + Cut} is positive but not decisive ($|\Delta\ln\mathcal{B}|=1.6$). However, in Appendix~\ref{appendix-3} we argue that the contribution from the aligned component in the \texttt{Aligned + Isotropic + Cut} needs to be small and gets more strongly disfavoured if very wide opening angles are excluded. Moreover, aligned models consistently disagree with the inferred excess fractions of the non-parametric model above, as summarised in Table~\ref{tab:models}. Thus, in what follows  we focus on the \texttt{Gaussian + Isotropic + Cut} model. Other models are discussed at the end of this section and in Appendix~\ref{appendix-3}. 

In Figure~\ref{fig:PPD-tilts}, we show the posterior predictive distribution (PPD) of black hole spin-orbit tilts. In the left panel, we only show the low-mass Gaussian component ($m_1\lesssim\tilde{m}$), i.e., we construct quantiles of
\begin{equation}
    \pi_{\rm G}(\cos\theta)=\pi_t(\cos\theta|\mu_{t,k},\sigma_{t,k}),
\end{equation}
where $k=1,2,\dots$ correspond to individual samples of the posterior obtained with \texttt{GWpopulation} (some of which are shown in Figure~\ref{fig:PPD-tilts} by thin blue lines). Thus, the PPD is a one-dimensional distribution on the black hole spin tilts which is obtained by marginalising over the two-dimensional distribution of $\cos\theta_{1(2)}$ in Equation~\eqref{eq:pi}. In the right panel, we include the low-mass isotropic component, i.e., we consider
\begin{equation}
    \pi_{\rm GI}(\cos\theta)=\xi_k\pi_t(\cos\theta|\mu_{t,k},\sigma_{t,k})+\frac{1-\xi_k}{2}.
\end{equation}
The right panel shows excellent agreement with the non-parametric B-Spline model of \citet{GWTC4pop,GWTC4pop-zenodo} \new{and earlier parametric analyses of \citet{Vitale2022},} which exhibit a similar peak near $\cos\theta\approx0$ and lends further credibility to our analysis. The left panel shows that its Gaussian component, which is responsible for the overabundance at near-perpendicular orientations, agrees well with the expected tilt distribution (red) of mergers caused by the Lidov--Kozai effect in triples \citep{Antonini2018}. However, it is in stark contradiction with expectations from traditional isolated binary formation scenario (orange) which peak sharply at $\cos\theta=1$ \citep{Olejak2024} even if large natal kicks are considered (dashed). We further discuss the astrophysical implications of our findings in Section~\ref{sec:astro}.

In Appendix~\ref{appendix-2}, we show that the PPD of the black hole spin tilt distribution of the low-mass Gaussian component is the result of posterior distributions with a small non-zero mean $\mu_t=0.20^{+0.21}_{-0.11}$ and standard deviation $\sigma_t=0.55^{+0.25}_{-0.16}$ which deviate significantly from their priors $\mu_t\sim\mathcal{U}_{[-1,1]}$ and $\sigma_t\sim\mathcal{N}_{[0.1,4]}(\mu=0,\sigma=1/2)$, respectively. We also find that the Gaussian component is the dominant component in the low-mass population ($\xi=0.86^{+0.11}_{-0.54}$) where the small isotropic contribution ($1-\xi$) causes the vertical shift of the PPD from the left to the right panel of Figure~\ref{fig:PPD-tilts}. \new{A complete absence of the Gaussian component  ($\xi=0$) is disfavoured but cannot be ruled out with certainty.} For the mass threshold that separates the low-mass population from the high-mass isotropic population we infer $\tilde{m}=44.26^{+8.60}_{-4.56}\,\rm M_\odot$ which agrees with \new{previous findings \citep{2022ApJ...941L..39W,2024PhRvL.133e1401L,2025PhRvL.134a1401A,Antonini2025,Tong2025,Plunkett2026}}.
%based on a broadening of the $\chi_{\rm eff}$ distribution and \citet{Tong2025} on the sparsity of secondary black hole masses above $\tilde{m}$.  

\begin{figure}
    \centering
    \includegraphics[width=\linewidth]{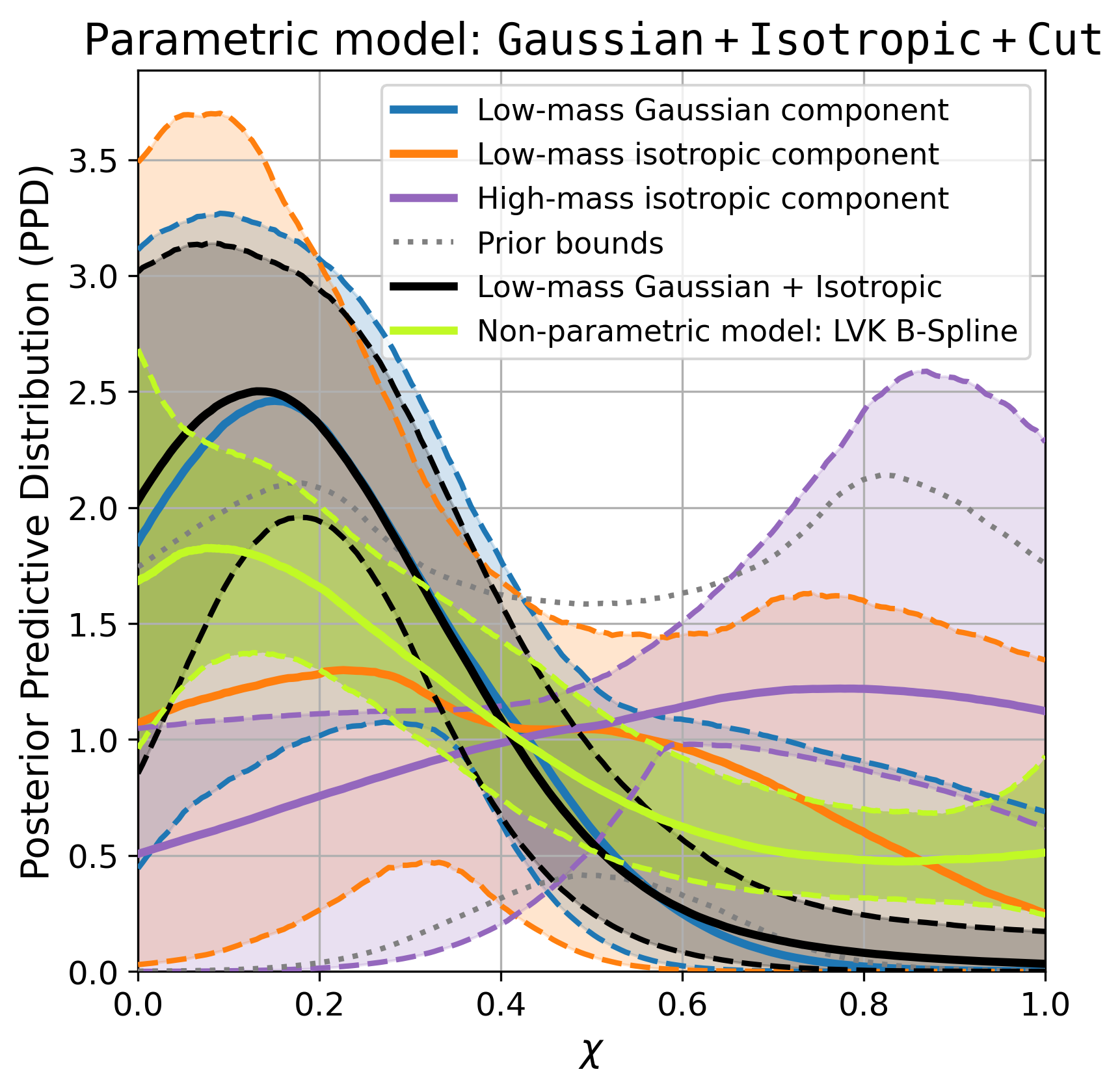}
    \caption{PPD of the black hole spin magnitude distribution in the \texttt{Gaussian Isotropic Cut} model. Blue shows the magnitude distribution of the low-mass Gaussian component (parametrised by $\mu_\chi$ and $\sigma_{\chi}$), orange of the low-mass isotropic component ($\mu^{\rm Iso}_\chi$ and $\sigma^{\rm Iso}_{\chi}$), and purple of the high-mass isotropic component ($\mu^{\rm HighIso}_\chi$ and $\sigma^{\rm HighIso}_{\chi}$). The transition between low- and high-mass components is inferred at a primary mass cut-off of $\tilde{m}\approx 44.3^{+8.7}_{-4.6}\,\rm M_\odot$. The green colour shows the non-parametric B-Spline model of \citet{GWTC4pop,GWTC4pop-zenodo}. Solid lines and shaded envelopes indicate medians and $90\,\%$ credible intervals, respectively. \new{The dotted lines show the prior bounds of the single-component models.}}
    \label{fig:PPD-mag}
\end{figure}

In Figure~\ref{fig:PPD-mag}, we present the PPDs for the spin magnitudes of the low-mass Gaussian component (parametrised by $\mu_\chi$ and $\sigma_{\chi}$), low-mass isotropic component ($\mu^{\rm Iso}_\chi$ and $\sigma^{\rm Iso}_{\chi}$), and high-mass isotropic component ($\mu^{\rm HighIso}_\chi$ and $\sigma^{\rm HighIso}_{\chi}$). The dominant low-mass Gaussian component (blue) tends to have small non-zero spins similar to findings of \citet{GWTC4pop}. The subdominant low-mass isotropic component (orange) tends to exhibit somewhat larger spin magnitudes. However, the large uncertainty hinders reliable interpretations. Meanwhile, the high-mass isotropic component (purple) appears to peak at significantly larger values $\gtrsim0.5$ which would be expected if those include highly spinning black holes formed through hierarchical mergers \citep{GerosaFishbach} or through gas-accretion in active galactic nuclei. We highlight that the mismatches of each component with the non-parametric B-Spline model by the \citet{GWTC4pop,GWTC4pop-zenodo} is expected as the latter fits for the entire merger population and does not differentiate between subpopulations.

In Appendix~\ref{appendix-3}, we present the mixing fractions $\xi$ of the Gaussian component across different models (except \texttt{Gaussian} and \texttt{Aligned} where no mixing was assumed, cf.~Table~\ref{tab:models}). The \texttt{Aligned + Isotropic} and \texttt{Aligned + Isotropic + Cut} models show that if alignment was enforced it would only contribute by $\xi=0.15^{+0.11}_{-0.06}$ and $0.09^{+0.66}_{-0.06}$, respectively, which further disfavours significant contribution from an aligned component. Comparing the \texttt{Gaussian + Isotropic} and \texttt{Uniform Gaussian + Isotropic} models to their \texttt{LVK} counterparts (which enforce the same spin magnitude distribution in the Gaussian and isotropic components) we observe a peculiar bi-modality around $\xi\approx0.2$ and 0.8 in the former. However, the peak around 0.2 vanishes if only the mixing within the low-mass population in the \texttt{Gaussian + Isotropic + Cut} model is considered, which suggests that it is an artefact of the isotropic high-mass population.

In Appendix~\ref{appendix-3} (Figure~\ref{fig:other-models}), we also show the PPDs of the spin-orbit tilts and spin magnitudes across all other models than \texttt{Gaussian + Isotropic + Cut}. In particular, it shows that models which enforce alignment of the Gaussian component (\texttt{Aligned}, \texttt{Aligned + Isotropic}, and \texttt{Aligned + Isotropic + Cut}) and \texttt{Uniform Gaussian + Isotropic} with a uniform prior on $\sigma_t$ fail to recover the peaked shape of the non-parametric B-Spline model \citep{GWTC4pop,GWTC4pop-zenodo}.

%% Please use the acknowledgment and contribution environments. This will 
%% be anonomyized when the "anonymous" style option is used. 

\section{Astrophysical implications}\label{sec:astro}

Most massive progenitor stars of black holes or neutron stars are found in hierarchical triples or higher-order configurations \citep[e.g.,][]{Moe2017,Offner2023}, where a close inner binary is orbited by one or more outer distant companions. The gravitational three-body dynamics of hierarchical triples can naturally lead to binary mergers in which the component spins are \new{preferentially} nearly perpendicular to the inner orbital angular momentum \citep{Antonini2018,Rodriguez2018}. \cite{Liu2018} showed that this arises
because the spin of a black hole in the inner binary evolves through a combination of de~Sitter precession around the inner orbital angular momentum $\boldsymbol{L}\equiv\boldsymbol{L}_\mathrm{in}$, the precession of $\boldsymbol{L}_\mathrm{in}$ itself around the outer orbital axis $\boldsymbol{L}_\mathrm{out}$, and the gradual shrinking of the inner binary orbit due to gravitational-wave energy-loss.

The evolution of the spin can be described by the precession equation $d\boldsymbol{\chi}/dt = \boldsymbol{\Omega}_{\rm p} \times \boldsymbol{\chi}$, where the effective precession (rotation) vector is given by $\boldsymbol{\Omega}_{\rm p} = \Omega_{\rm L}\,\boldsymbol{\hat{L}}_\mathrm{out} + \Omega_{\rm SL}\,\boldsymbol{\hat{L}}_\mathrm{in}$. Here, $\Omega_{\rm L}$ and $\Omega_{\rm SL}$ denote the precession rates associated with the orbital motion of the inner binary around the tertiary and the de~Sitter spin--orbit coupling, respectively. The direction of $\boldsymbol{\Omega}_{\rm p}$ defines the instantaneous precession axis of the spins.

In the adiabatic regime, where the evolution of $\boldsymbol{\Omega}_{\rm p}$ is slow compared to the spin precession rate, the quantity $\cos\bar{\theta}_{\rm p} = \boldsymbol{\hat{\chi}}\!\cdot\!\boldsymbol{\hat{\Omega}}_{\rm p}$ is nearly conserved, implying that the angle between the spin vector and the precession axis remains approximately constant. At large separations, $\boldsymbol{\Omega}_{\rm p}$ is dominated by the outer orbital term, such that $\boldsymbol{\Omega}_{\rm p} \simeq \Omega_{\rm L}\,\boldsymbol{\hat{L}}_\mathrm{out}$. If the component spins are initially aligned with $\boldsymbol{L}_\mathrm{in}$, and $\boldsymbol{L}_\mathrm{in}$ is inclined close to $90^{\circ}$ relative to $\boldsymbol{L}_\mathrm{out}$ as required for Lidov--Kozai mergers, the spins are initially nearly perpendicular to $\boldsymbol{\Omega}_{\rm p}$ (i.e. $\theta_{\rm p} \approx 90^{\circ}$). 

As the tertiary companion drives the inner binary to large eccentricities where efficient gravitational-wave emission at close pericentre passages shrinks the inner orbit, the coupling to the tertiary weakens, and $\Omega_{\rm SL}$ gradually overtakes $\Omega_{\rm L}$, causing $\boldsymbol{\hat{\Omega}}_{\rm p}$ to transition smoothly from being aligned with $\boldsymbol{L}_\mathrm{out}$ to being aligned with $\boldsymbol{L}_\mathrm{in}$. Because $\theta_{\rm p}$ is an adiabatic invariant, the spin maintains its inclination of $\simeq 90^{\circ}$ relative to the evolving precession axis. By the time the system decouples and the inner binary merges, the component spins therefore lie nearly within the orbital plane, leading to small values of $\chi_{\mathrm{eff}}$ but potentially large in-plane spin components. Meanwhile, the angles between the two spins in the orbital plane of the binary may be distributed across the whole range between 0 and $180^{\circ}$ due to generally different spin precession rates of the black holes if $m_1\ne m_2$ \citep[][see Figure 10 therein]{Rodriguez2018}.

In contrast, binary black hole mergers formed through the evolution of isolated binary stars are generally expected to have spins closely aligned with the orbital angular momentum, resulting in low misalignment angles and a preference for positive effective spins \citep{Kalogera2000,Farr2017,2018PhRvD..98h4036G,Belczynski2020,Bavera2020,OlejakSpins2021,Broekgaarden2022}. In standard binary-evolution models, stellar spins are assumed to be aligned with the orbital angular momentum prior to core collapse, and any natal kick imparted during the collapse tilts the orbital plane, thereby directly setting the resulting spin–orbit misalignment \citep[e.g.,][]{Brandt1995, Kalogera2000,Fragos2010,Tauris2017}, with the maximum tilt limited by the risk to disrupt the binary. As exemplified in Figure~\ref{fig:PPD-tilts}, even under the assumption of very high black hole natal kicks—which are not expected for massive black holes from an evolutionary perspective \citep{Janka2024}—the resulting distribution remains strongly peaked at modest tilt angles, corresponding to $\cos\theta \gtrsim 0.75$.

Producing a distribution with a significant fraction of systems at $\cos\theta\lesssim0.75$ via isolated binary channels requires non-standard and poorly constrained assumptions, such as highly misaligned progenitor spins,  ``spin tossing" at black hole formation \citep{Tauris2022}, or finely tuned correlations between spin orientations and natal-kick directions \citep{Baibhav2024}. In particular, reproducing an observed peak near $\cos\theta \approx 0$ would require preferentially orienting black hole spins perpendicular to the natal kicks \citep[][see Figure 5 therein]{Baibhav2024}. Such an assumption is highly non-standard and is challenged by observational constraints from, e.g., pulsars' velocity-spin alignment \citep{Johnston2005,Noutsos2012,Noutsos2013,Mandel2023} as well as by simulation results of core-collapse supernovae \citep[e.g.,][]{Burrows2024}.

Another possibility discussed in the literature is a spin flip induced by mass transfer \citep{StegmannAntonini}, which could produce in-plane tilts. This mechanism could reduce the first-born black hole’s spin projection onto the orbital angular momentum; however, its nature and efficiency remain highly uncertain as it requires very efficient angular momentum transport between the stellar core and envelope and inefficient tides which would otherwise realign the spins after mass transfer.

Alternatively, \citet{VitaleMould2025} discuss the possibility that parametric modelling for an underlying preferentially aligned merger population may lead to inference of spurious peaks away from perfect alignment due to measurement uncertainties and the finiteness of the current sample size. \new{This highlights that} it is essential to verify the non-parametric inference of \citet{GWTC4pop} and our parametric analysis with the growing gravitational-wave dataset in the future. However, we stress that \citet{VitaleMould2025} tested underlying preferentially aligned populations ($\mu_t=1$) which are extremely broad with $\sigma_t=1.15$, which seems inconsistent with current expectations from astrophysical models that tend to be narrower (see Section~\ref{sec:methods}). \new{Even in that case, their spurious peaks away from alignment are not as narrowly peaked as close to $\cos\theta\approx0$ as the feature in the GWTC-4.0 data.}

\section{Summary and discussion}\label{sec:summary}
The latest gravitational-wave data GWTC-4.0 has enabled \citet{GWTC4pop} to infer a global peak near $\cos\theta\approx0$ in non-parametric models for the spin-orbit tilt angle distribution of the binary black hole merger population. In this work, we have recovered this distribution with parametric models that contain a dominant pronounced Gaussian peak at near-perpendicular directions. This defies traditional formation scenarios from isolated massive binary stars, which recover significant spin-orbit misalignment only under fine-tuned or highly uncertain assumptions of the binary evolution and stellar collapse (Section~\ref{sec:astro}). Instead, the latest gravitational-wave data suggests an alternative explanation for the main origin of binary black hole mergers, which builds upon the fact that most massive black hole progenitor stars are found in hierarchical triple or higher-order configurations \citep{Moe2017,Offner2023}. In these systems, the relativistic gravitational dynamics of three-body systems naturally produces mergers with near-perpendicular spin angles \citep{Antonini2018,Liu2018}, \new{while they could also yield event rates \citep[][and references therein]{MandelBroekgaarden2022} which are broadly consistent with gravitational-wave observations \citep{GWTC4pop}. Triples have also been shown to be in overall agreement with key features of the observationally inferred mass distributions such as a global peak at $\sim10\,\rm M_\odot$ and declining rate towards higher masses \citep[e.g.,][]{Antonini2017,Silsbee2017,Stegmann2022,Stegmann2022b}. However, we note that there are no comprehensive studies that investigate the impact of assumptions in the stellar evolution (e.g., inner binary mass transfer or natal kicks) and account for delay time distribution and metallicity-dependent star-formation history, which are needed for robust comparison to gravitational-wave data.}

We highlight that the spin distribution from triples also agrees qualitatively well with the inferred distribution of $\chi_{\rm eff}$, which is skewed and
asymmetric about zero with more support for
positive values \citep{GWTC4pop,Banagiri2025b}. Due to the natural preference to produce tilts at $\cos\theta\approx0$ similar $\chi_{\rm eff}$ distributions have been obtained for triples, largely independent of the assumed spin magnitudes \citep[e.g.,][see Figure~9 therein]{Rodriguez2018}. In addition, a triple formation scenario could contribute to the growing number of claims about mergers with residual orbital eccentricity \citep{Romero-Shaw2022,Gupte2024,2024ApJ...972...65I,2025arXiv250415833D,Morras2025,Phukon2025}, which would be impossible to recover from isolated binary star evolution \citep[e.g.,][]{Belczynski2002,Fumagalli2024}. While some of the eccentric binary black hole candidates have masses that are confidently within the upper mass gap and may only plausibly explained by hierarchical mergers in dense environments \citep[e.g.,][]{GW190521,Gayathri2020}, candidates below the upper mass could be also explained by the Lidov--Kozai triples mechanism \citep{2025PhRvD.112f3052R,McMillin2025}, \new{where the subset of non-adiabatic triples is expected to yield a small fraction of eccentric outlier detections \citep[e.g.,][]{Antonini2017,Rodriguez2018,Liu2019b}}. In particular, it has been argued that recent claims for residual eccentricity in one neutron star-black hole (NSBH) merger \citep{Morras2025,planas2025,Kacanja2025,Jan2025} may only be obtained through triple star evolution \citep{Stegmann2025} and imply a dominant contribution to the total NSBH merger rate \citep{RomeroShawStegmannMorrasEccNSBH_inprep}.

If black hole mergers indeed tend to result from triple rather than binary star evolution, it would suggest that binary mass transfer is less efficient in forming very close binary black hole systems than previously assumed, e.g., in scenarios involving successful common envelope ejection \citep{Belczynski2020}. This may indicate that binary black hole progenitors instead tend to undergo stable mass transfer \citep{Gallegos-Garcia2021}, which typically produces systems with wider orbital separations than those expected from a common envelope scenario \citep{Bavera2020, Olejak2021, vanSon2022}. Moreover, recent studies have predicted that the minimum separation of binary black hole systems formed via binary evolution may be limited by a delayed unstable mass transfer and stellar mergers \citep{Klencki2025}. Also, observations of other types of binaries hosting compact objects further challenge standard binary evolution models. Notable examples are Gaia black holes \citep{ElBadry2023a, Chakrabarti2023, ElBadry2023b}, whose properties cannot be reconciled with conventional models \citep{Nagarajan2025} and likely require non-standard assumptions about angular momentum loss during mass transfer \citep{Olejak2025}. %In this context, evolution within a triple system could play a critical role, enabling inner binary systems that would otherwise be too wide after mass transfer to merge on much shorter timescales.

%% To help institutions obtain information on the effectiveness of their 
%% telescopes the AAS Journals has created a group of keywords for telescope 
%% facilities.
%
%% Following the acknowledgments section, use the following syntax and the
%% \facility{} or \facilities{} macros to list the keywords of facilities used 
%% in the research for the paper.  Each keyword is check against the master 
%% list during copy editing.  Individual instruments can be provided in 
%% parentheses, after the keyword, but they are not verified.

\section*{Acknowledgments}
\neww{We thank the two anonymous reviewers, Colm Talbot, Isobel Romero-Shaw, and Carlos Lousto for useful comments and input that helped improving this work.} 
FA and VR are supported by the UK’s Science and Technology Facilities Council
grants ST/V005618/1 and UKRI2489. 
S.R.~is funded by the Deut\-sche For\-schungs\-ge\-mein\-schaft (DFG, German Research Foundation) – project number 546677095. This material is based upon work supported by NSF's LIGO Laboratory which is a major facility fully funded by the National Science Foundation.

%% Similar to \facility{}, there is the optional \software command to allow 
%% authors a place to specify which programs were used during the creation of 
%% the manuscript. Authors should list each code and include either a
%% citation or url to the code inside ()s when available.
\software{\texttt{GWpopulation} \citep{Talbot2019,Talbot2025}; \texttt{Bilby} \citep{Bilby}; \texttt{GWpopulation Pipe} \citep{gwpop_pipe}}

\appendix

\twocolumngrid

\section{\new{Mass and redshift modelling}}\label{appendix-1}
\new{In all our models, we assume a simple parameterisation of our mass distribution as \citep{Talbot2019,Talbot2025}}
\begin{equation}
    \pi(m_1, q) = \pi(m_1) \pi(q | m_1),
\end{equation}
\new{where $0<q\le1$ is the binary mass ratio and}
\begin{align}
    \pi(m_1) &= (1 - \lambda_m) \pi_{\text{pow}}(m_1) + \lambda_m \pi_{\text{norm}}(m_1),\label{eq:mass}\\
    \pi_{\text{pow}}(m_1) &\propto m^{-\alpha} \quad\text{with}\quad m_{\min} \leq m_1 < m_{\max},\\
    \pi_{\text{norm}}(m_1) &\propto \exp\left(-\frac{(m_1 - \mu_{m})^2}{2\sigma^2_m}\right) \nonumber\\
    &\quad\text{with}\quad m_{\min} \leq m_1 < 100\,\rm M_\odot,\\
    \pi(q|m_1) &= \frac{1 + \beta}{1 - (m_{\min}/m_1)^{1 + \beta}} q^\beta.
\end{align}
\new{Thus, the primary mass distribution is modelled as a power-law with a Gaussian component and the conditional mass-ratio distribution as a power-law.}

\new{For the distribution in redshift $z$ we assume a power-law following \cite{2018ApJ...863L..41F}}

\begin{equation}\label{eq:redshift}
    \pi(z) = (1 + z)^\lambda.
\end{equation}

\new{For all hyper-parameters introduced in Equations~\eqref{eq:mass}~--~\eqref{eq:redshift} we assume uniform priors within the bounds given in Table~\ref{tab:mz_priors}~\citep{Talbot2019,Talbot2025}}.

\begin{table}[t]
\centering
\begin{tabular*}{\columnwidth}{@{\extracolsep{\fill}} l c c}
\hline
Parameter & Description & Prior \\
\hline
$\lambda_m$ & $m_{1}$ mixture fraction & $\mathcal{U}_{[0,1]}$ \\
$\alpha$ & $m_{1}$ power-law index & $\mathcal{U}_{[-2,4]}$ \\
$m_{\min}/\mathrm{M}_\odot$ & minimum mass & $\mathcal{U}_{[2.0,2.5]}$ \\
$m_{\max}/\mathrm{M}_\odot$ & maximum mass & $\mathcal{U}_{[80,100]}$ \\
$\mu_m/\mathrm{M}_\odot$ & $m_{1}$ Gaussian component peak & $\mathcal{U}_{[10,50]}$ \\
$\sigma_m/\mathrm{M}_\odot$ & $m_{1}$ Gaussian component width & $\mathcal{U}_{[1,10]}$ \\
$\beta$ & mass ratio power-law index & $\mathcal{U}_{[-4,12]}$ \\
$\lambda$ & redshift power-law index & $\mathcal{U}_{[-1,10]}$ \\
\hline
\end{tabular*}
\caption{Prior distributions for mass and redshift model hyper-parameters.}
\label{tab:mz_priors}
\end{table}

\new{We emphasise that for models that mix two components (3~--~9 in Table~\ref{tab:models}) we assume the \textit{same} mass and redshift modelling for both, which is consistent with the approach of the LVK collaborations \citep[e.g.,][]{2023PhRvX..13a1048A,GWTC4pop}. However, if both components represent two different astrophysical formation channels (e.g., a near-perpendicular Gaussian from triple star evolution and an isotropic component from dynamical interactions in star clusters) they could in reality follow different mass and redshift distributions \citep[cf.,][]{Dorozsmai2025}. Investigating the impact of our simplification as well as allowing for any spin correlation with mass or redshift are left for future work.}

\section{Supplementary results about the \texttt{Gaussian + Isotropic + Cut} model}\label{appendix-2}

\new{In Figure~\ref{fig:long_custom_corner} we show a corner plot of all hyper-parameters of the \texttt{Gaussian + Isotropic + Cut} model. Regarding the spin properties, it shows that the near-perpendicular Gaussian component ($\mu_t=0.20^{+0.21}_{-0.11}$ and $\sigma_t=0.55^{+0.25}_{-0.16}$) is the dominant component ($\xi=0.86^{+0.11}_{-0.54}$) at low masses ($m_1\lesssim\tilde{m}=44.26^{+8.60}_{-4.56}\,\rm M_\odot$). In addition, Figure~\ref{fig:zeta_m1} displays that the PPD of the $\zeta(m_1,\tilde{m})$ for posterior draws of $\tilde{m}$, showing that the transition is significantly different from the prior draws.} \neww{Moreover, Figure~\ref{fig:long_custom_corner_variable_k} shows the corner plot of all hyper-parameters if $\tilde{k}$ in Equation~\eqref{eq:zeta} is treated as a free parameter (see Section~\ref{sec:methods}). Comparison to Figure~\ref{fig:long_custom_corner} shows that the posterior distribution of all other parameters remain largely similar. We also find that the \texttt{Gaussian + Isotropic + Cut} model remains preferred over \texttt{Aligned + Isotropic + Cut} model if $\tilde{k}$ is treated as a free parameter in both cases with $\Delta\ln\mathcal{B}=-2.5$.}

\begin{comment}
In Figure~\ref{fig:corner-spins} we show a corner plot of selected parameters of the \texttt{Gaussian + Isotropic + Cut} model, showing that the near-perpendicular Gaussian component ($\mu_t=0.20^{+0.21}_{-0.11}$ and $\sigma_t=0.55^{+0.25}_{-0.16}$) is the dominant component ($\xi=0.86^{+0.11}_{-0.54}$) at low masses ($m_1\lesssim\tilde{m}=44.26^{+8.65}_{-4.57}\,\rm M_\odot$). \new{For completeness, Figure~\ref{fig:long_custom_corner} shows the full corner plot including the hyper-parameters describing the mass and redshift modelling in Appendix~\ref{appendix-1}.}

\begin{figure}
    \centering
    \includegraphics[width=\linewidth]{custom_corner.png}
    \caption{Joint and marginal posteriors for the mean $\mu_t$ and standard deviation $\sigma_t$ of the Gaussian component, its mixing fraction $\xi$, and the mass cut-off $\tilde{m}$ in the \texttt{Gaussian + Isotropic + Cut} model. \new{The full corner plot of this model is shown in Figure~\ref{fig:long_custom_corner}.}}
    \label{fig:corner-spins}
\end{figure}
\end{comment}

\begin{figure*}
    \centering
    \includegraphics[width=\linewidth]{long_custom_corner.png}
    \caption{\new{All joint and marginal posteriors in the \texttt{Gaussian + Isotropic + Cut} model. Variables are defined in Section~\ref{sec:methods} and Appendix~\ref{appendix-1}.}}
    \label{fig:long_custom_corner}
\end{figure*}

\begin{figure}
    \centering
    \includegraphics[width=\linewidth]{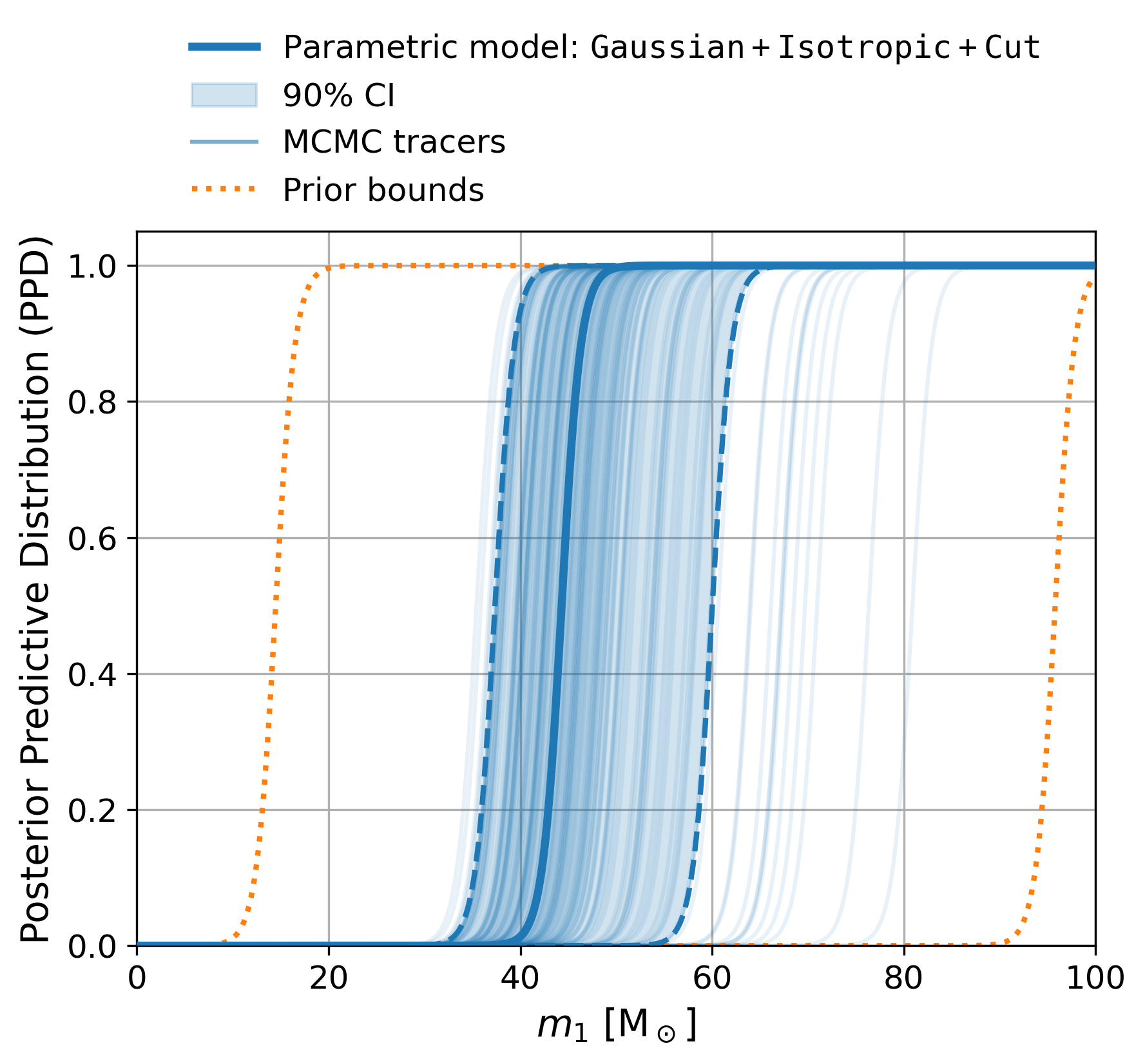}
    \caption{\new{PPD of the cut-off Equation~\eqref{eq:zeta} in the \texttt{Gaussian + Isotropic + Cut} model. Blue shows the PPD for posterior draws of $\tilde{m}$, and orange shows the $90\%$ bounds for prior draws of $\tilde{m}$ (uniform distribution $\mathcal{U}_{[10\,\rm M_\odot,100\,M\odot]}$, see Section~\ref{sec:methods}).}}
    \label{fig:zeta_m1}
\end{figure}

\begin{figure*}
    \centering
    \includegraphics[width=\linewidth]{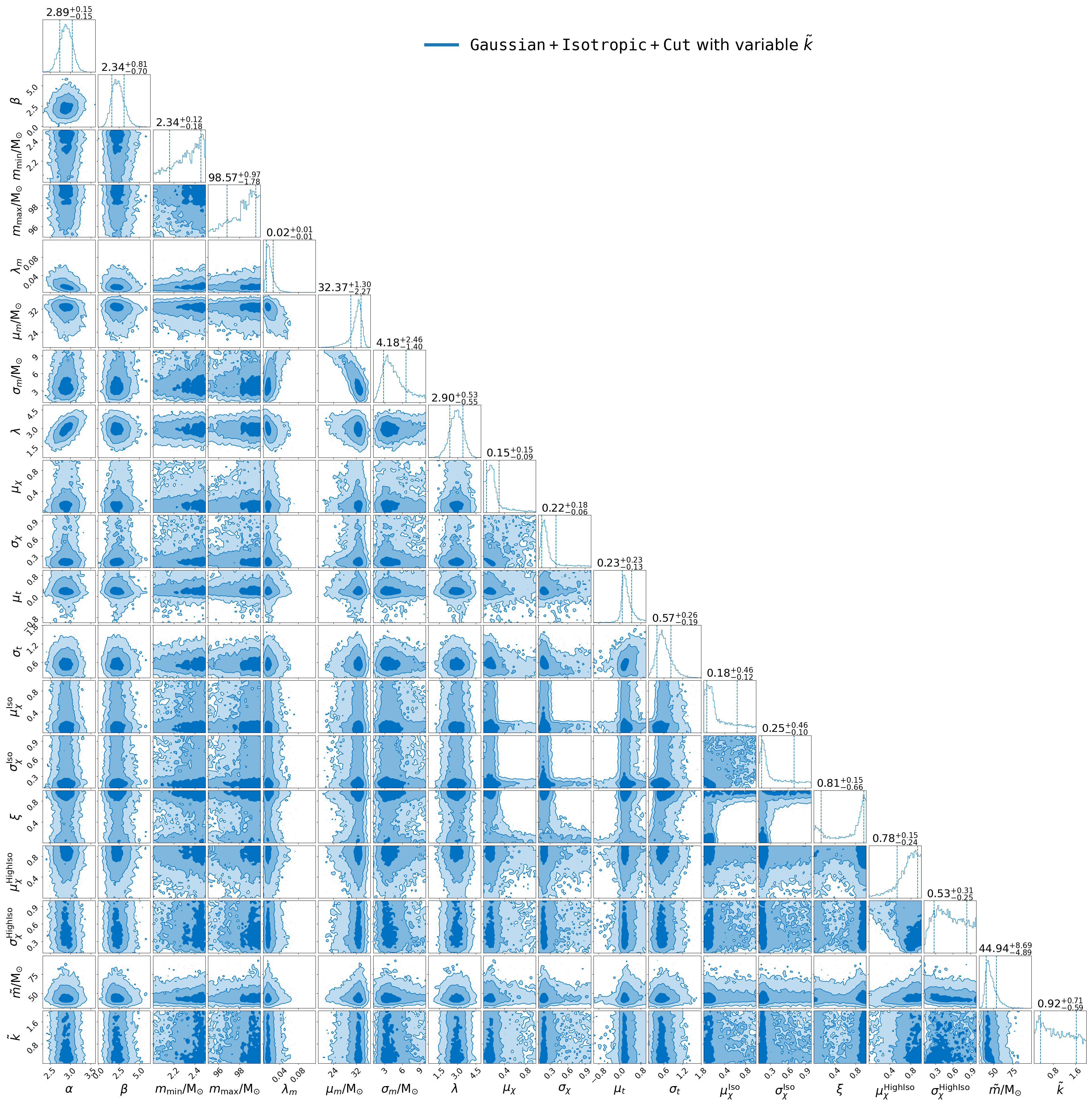}
    \caption{\neww{Same as Figure~\ref{fig:long_custom_corner} with a free parameter $\tilde{k}$.}}
    \label{fig:long_custom_corner_variable_k}
\end{figure*}

\section{Other population models}\label{appendix-3}
\begin{figure*}
    \centering
    \includegraphics[width=\linewidth]{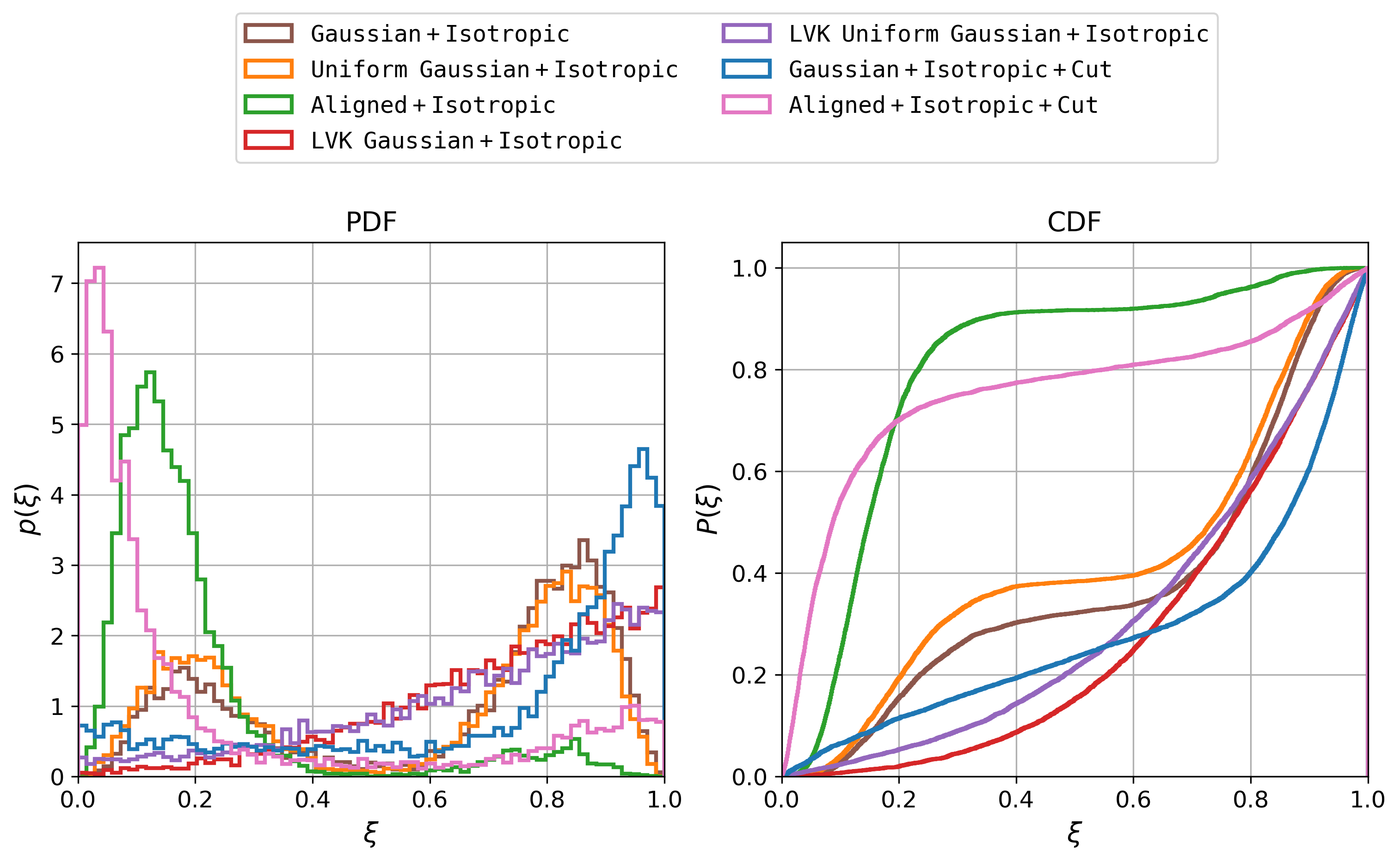}
    \caption{Marginal posterior distributions of the Gaussian mixing fraction $\xi$. The left panel shows the probability density function (PDF); the right panels its cumulative density function (CDF). The fraction $\xi$ refers to the mixing fraction of the Gaussian component, which is enforced at alignment in the \texttt{Aligned + Isotropic} model ($\mu_t=1$) or moves freely in all others ($-1\le\mu_t\le1$), whereas $1-\xi$ refers to the isotropic fraction. For the \texttt{Gaussian + Isotropic + Cut} and \texttt{Aligned + Isotropic + Cut} models the fractions $\xi$ and $1-\xi$ describe the mixing in the low-mass population ($m_1\lesssim\tilde{m}$).}
    \label{fig:mixing-fractions}
\end{figure*}
\begin{figure}
    \centering
    \includegraphics[width=\linewidth]{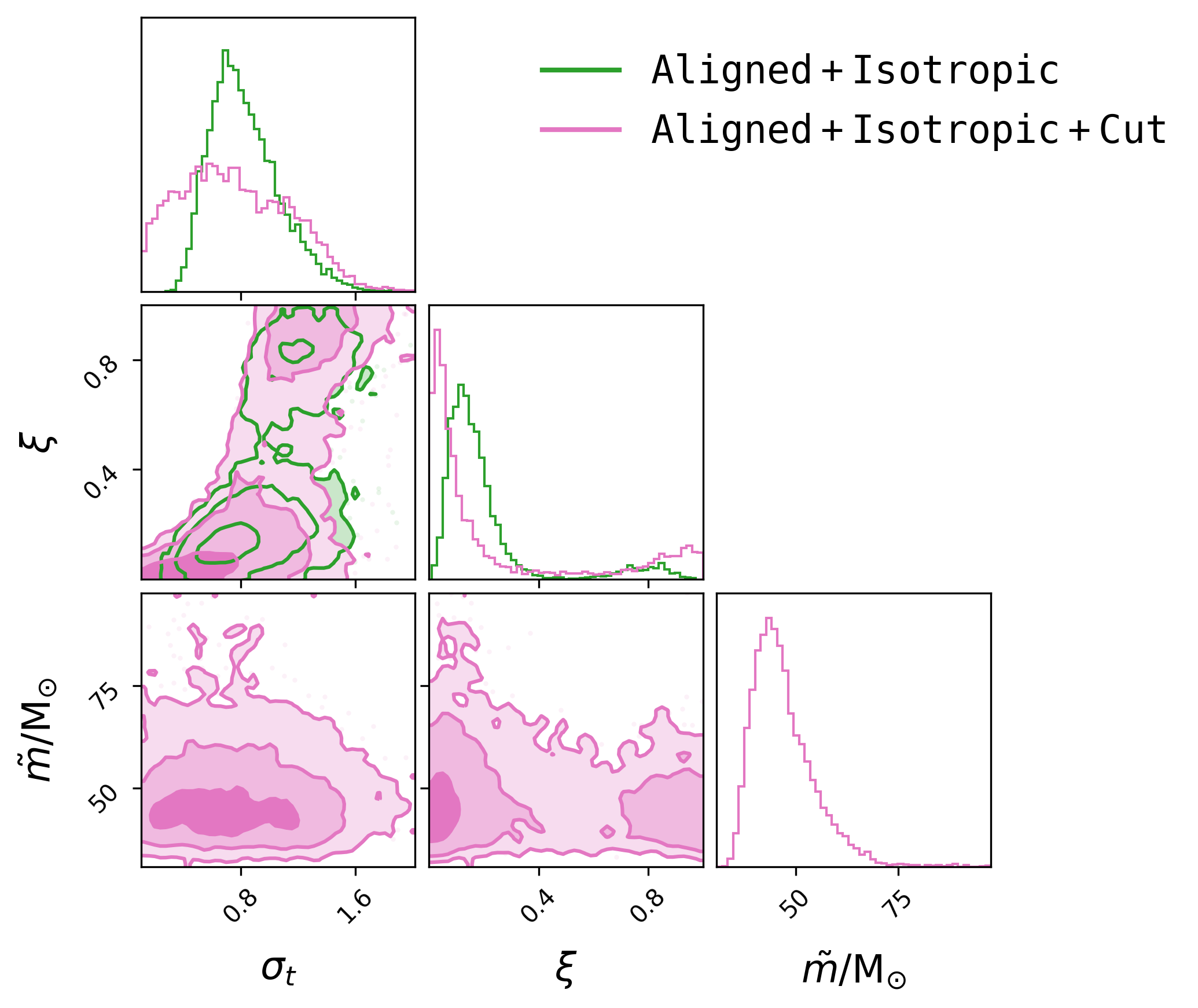}
    \caption{Joint and marginal posteriors for the standard deviation $\sigma_t$ of the aligned component ($\mu_t=1$) and its mixing fraction $\xi$ in the \texttt{Aligned + Isotropic} model (green) as well as the mass cut-off $\tilde{m}$ in the \texttt{Aligned + Isotropic + Cut} model (pink).}
    \label{fig:Aligned_Isotropic_vs_Aligned_Isotropic_Cut_corner}
\end{figure}

Figure~\ref{fig:mixing-fractions} shows the marginal posterior distributions of the Gaussian mixing fraction $\xi$. In all models, where the Gaussian is allowed to move freely ($-1\le\mu_t\le1$) it dominates the distribution, while allowing for a potential mixing with a subdominant isotropic component. Enforcing a preferentially aligned spin-orbit configuration ($\mu_t=1$ in \texttt{Aligned + Isotropic} and \texttt{Aligned + Isotropic + Cut}) makes the Gaussian component a subdominant contribution, suggesting no preference for alignment in the gravitational-wave data. Figure~\ref{fig:mixing-fractions} shows only small probabilities for the aligned components to be dominant, e.g., the cumulative density function yields $P(\xi\lesssim0.8)\approx0.95$ and~$0.85$ in the \texttt{Aligned + Isotropic)} and \texttt{Aligned + Isotropic + Cut} models, respectively. However, Figure~\ref{fig:Aligned_Isotropic_vs_Aligned_Isotropic_Cut_corner} shows that a large aligned contribution $\xi\lesssim1$ is correlated with large width $\sigma_t\approx1$ (which is different to the \texttt{Gaussian + Isotropic + Cut} model in Figure~\ref{fig:long_custom_corner}). Thus, we conclude that the Bayes factors disfavour aligned models at varying degree (see Section~\ref{sec:results}) and even if realised they are either a subdominant component (Figure~\ref{fig:mixing-fractions}) or they dominate but are too wide to be plausibly explained by isolated binary formation scenarios (see Section~\ref{sec:astro}) and indistinguishable from the isotropic component (see Figure~\ref{fig:other-models} below).

In Figure~\ref{fig:rainbow}, we further investigate the preference against an aligned component. We explore variants where the aligned component is restricted to $\cos\theta\in[t_{\min},1]$ for some fixed threshold value $t_{\min}$, since realistic isolated binary evolution scenarios may yield spin-orbit tilts which are more narrowly confined than the entire range $\cos\theta\in[-1,1]$. For this purpose, we \new{follow a similar approach to \citet{Callister2022,Tong2022} and} adopt the \texttt{Aligned + Isotropic + Cut} model but use truncated normal distributions $\mathcal{N}_{[t_{\min},1]}(\cos\theta_i | \mu_t = 1, \sigma_t)$ in Equation~\eqref{eq:pi_cos} with $t_{\min}=-1.0,-0.9,-0.8,\dots,0.9$. The left panel of Figure~\ref{fig:rainbow} shows that larger values of $t_{\min}$ (i.e., more narrowly distributed spin-orbit angles about alignment) require smaller contributions from the aligned component, ranging from $\xi(t_{\min}=-1)=0.09^{+0.66}_{-0.06}$ to $\xi(t_{\min}=0.9)=0.01\pm0.01$. In the right panel, we compare the modified \texttt{Aligned + Isotropic + Cut} models to the \texttt{Gaussian + Isotropic + Cut} model (where the Gaussian component remains confined to $\cos\theta\in[-1,1]$). Evidence against the aligned models is positive \citep{Kass} with $|\Delta\ln\mathcal{B}|\approx1$ to 3 across the whole range of $t_{\min}$. We also might see a tendency that above $t_{\min}\gtrsim0.3$ models become more strongly disfavoured, which would also roughly coincide with the depletion of spin-orbit tilt angles in the non-parametric B-Spline model (cf. Figure~\ref{fig:PPD-tilts}). However, we do note that the scatter in $\Delta\ln\mathcal{B}$ across different explored models is rather large and inhibits definite conclusions. Since $\Delta\ln\mathcal{B}$ scales roughly with the detected sample size \citep{Kass} we expect the robustness of the model comparison to improve soon as the gravitational-wave catalogue grows during the \new{the fourth and fifth, as well as a possible intermediate, LVK observing runs.}

\begin{figure*}
    \centering
    \includegraphics[width=0.7\linewidth]{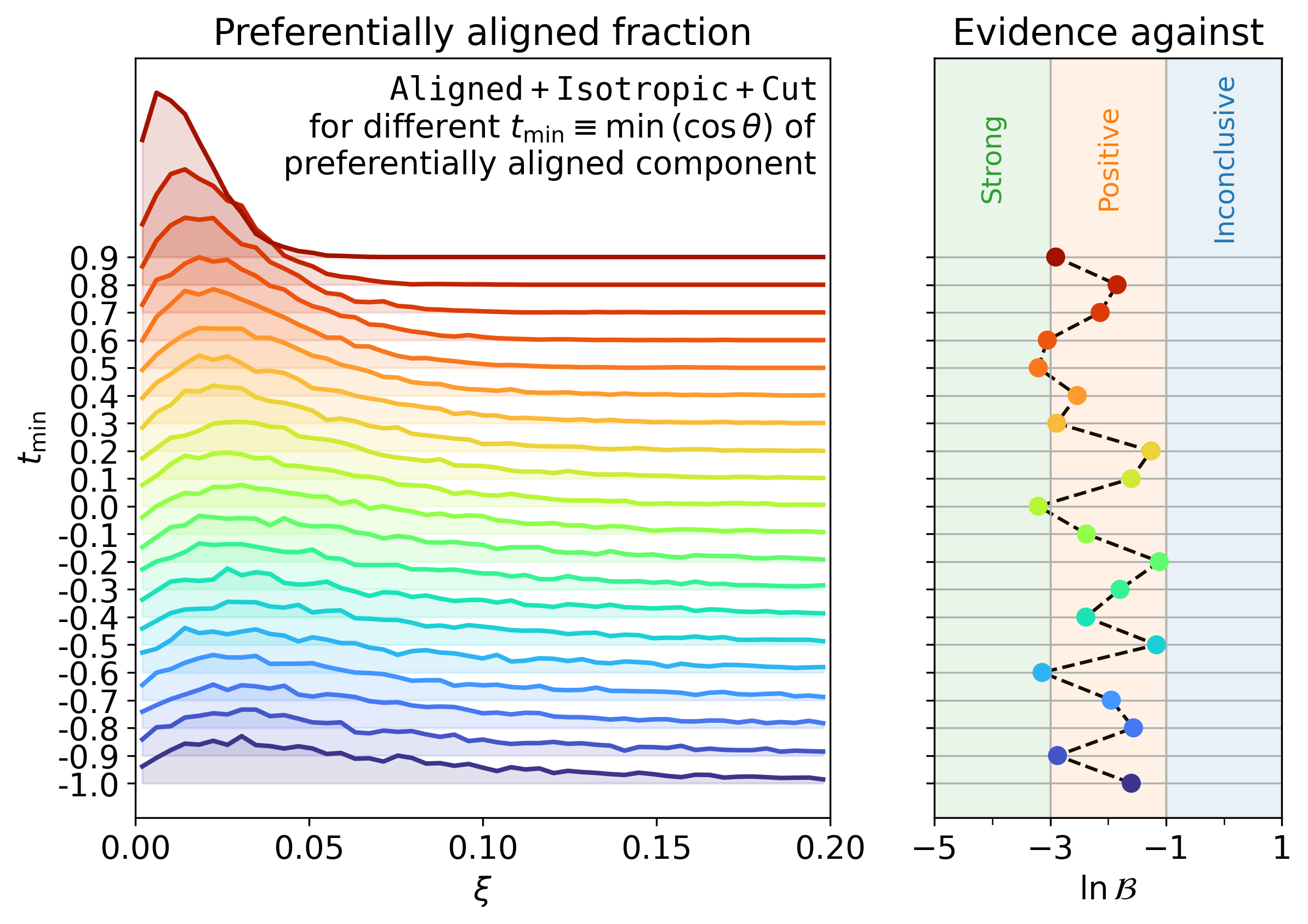}
    \caption{\texttt{Aligned + Isotropic + Cut} model for different lower bounds on $\cos\theta_i$ of the aligned component, i.e., $\mathcal{N}_{[t_{\min},1]}(\cos\theta_i | \mu_t = 1, \sigma_t)$ in Equation~\eqref{eq:pi_cos} with $t_{\min}=-1.0,-0.9,-0.8,\dots,0.9$. The left panel shows the marginal posterior distribution of the contribution $\xi$ of the low-mass aligned component (whereas $1-\xi$ corresponds to the low-mass isotropic component). The right panel shows the strength of evidence against each aligned model compared to our default \texttt{Gaussian + Isotropic + Cut} model, which allows but does not enforce preferentially aligned mergers. Nomenclature follows \citet{Kass}.}
    \label{fig:rainbow}
\end{figure*}

In addition, Figure~\ref{fig:other-models} shows that that the PPD of the spin-orbit tilts in most models with a free Gaussian recover the pronounced peak at $\cos\theta\approx0$ of the non-parametric model \citep{GWTC4pop}. The peak is less pronounced and recovered in models that assume a uniform prior on $\sigma_t$ (\texttt{Uniform Gaussian + Isotropic} and \texttt{LVK Uniform Gaussian + Isotropic}), where shallow distributions with large values of $\sigma_t$ mimic an isotropic distribution. As discussed in Section~\ref{sec:methods}, a uniform prior on $\sigma_t$ is not expected from astrophysical source modelling. We note that \texttt{LVK Uniform Gaussian + Isotropic} is a similar to the \texttt{Gaussian + Isotropic} model by \citet{GWTC4pop} and recovers a similar PPD.

\begin{figure*}
    \centering
    \includegraphics[width=\linewidth]{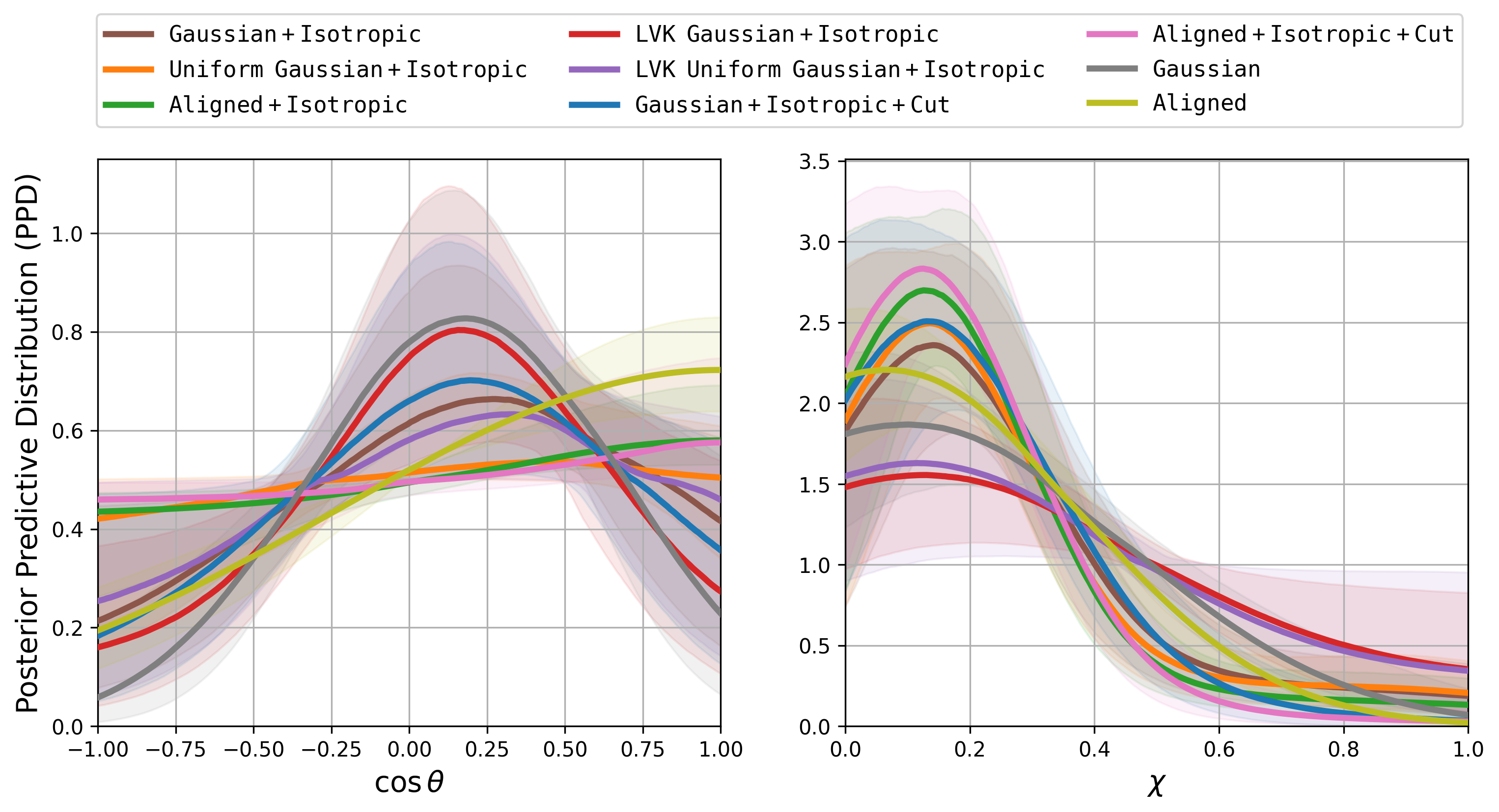}
    \caption{PPDs of the spin-orbit tilt angles (left panel) and spin magnitudes (right panel) across different models. In all models we show the Gaussian + Isotropic mixture (except for the single-component models \texttt{Gaussian} and \texttt{Aligned}).}
    \label{fig:other-models}
\end{figure*}

%% Appendix material should be preceded with a single \appendix command.
%% There should be a \section command for each appendix. Mark appendix
%% subsections with the same markup you use in the main body of the paper.
%%
%% Each Appendix (indicated with \section) will be lettered A, B, C, etc.
%% The equation counter will reset when it encounters the \appendix
%% command and will number appendix equations (A1), (A2), etc. The
%% Figure and Table counter will not reset.

%% For this sample we use BibTeX plus aasjournalv7.bst to generate the
%% the bibliography. The sample7.bib file was populated from ADS. To
%% get the citations to show in the compiled file do the following:
%%
%% pdflatex sample7.tex
%% bibtext sample7
%% pdflatex sample7.tex
%% pdflatex sample7.tex

\bibliography{sample701}{}
\bibliographystyle{aasjournalv7}

%% This command is needed to show the entire author+affiliation list when
%% the collaboration and author truncation commands are used.  It has to
%% go at the end of the manuscript.
%\allauthors

%% Include this line if you are using the \added, \replaced, \deleted
%% commands to see a summary list of all changes at the end of the article.
%\listofchanges

\end{document}